%% file: main.tex
\documentclass[a4paper, amsfonts, amssymb, amsmath, reprint, showkeys, nofootinbib, twoside,superscriptaddress]{revtex4-1}
\usepackage[english]{babel}
\usepackage[utf8]{inputenc}
\usepackage[colorinlistoftodos, color=green!40, prependcaption]{todonotes}
\usepackage{url}

\usepackage{color}
\input{preamble}

\input{auxcommands}

\usepackage[pdftitle={Article}, pdfauthor={Author}]{hyperref} 

\usepackage{comment}

\usepackage{graphicx}

\begin{document}
\title{Experimental characterization of quantum processes: a selective and efficient method in arbitrary finite dimension}

\author{Q. Pears Stefano}
    \email[Correspondence email address:]{email@institution.com}
    \affiliation{Departamento de Física, Facultad de Ciencias Exactas
      y Naturales, Universidad de Buenos Aires, 1428 Bueno Aires, Argentina}
    \affiliation{Consejo Nacional de Investigaciones Cient\'ificas y
      T\'ecnicas, 1425 Buenos Aires, Argentina}
    \author{I. Perito}
    \affiliation{Departamento de Física, Facultad de Ciencias Exactas
      y Naturales, Universidad de Buenos Aires, 1428 Bueno Aires, Argentina}
    \affiliation{Instituto de F\'iscia de Buenos Aires, CONICET, Universidad de Buenos Aires, Facultad de Ciencias Exactas y Naturales, Argentina}
\author{J. J. M. Varga}
    \affiliation{Centro de F\'isica de Materiales, Paseo Manuel de Lardizabal 5, 20018 Donostia-San Sebasti\'an, Spain.}
    \affiliation{Donostia International Physics Center, Paseo Manuel de Lardizabal 4, 20018 Donostia-San Sebasti\'an, Spain.}
\author{L. Reb\'on}
    \affiliation{Instituto de Física de La Plata, CCT La Plata,
      CONICET and Departamento de Física, Facultad de Ciencias
      Exactas, Universidad de La Plata, Diag. 113 e/63 y 64 La Plata, Argentina}
    \affiliation{Departamento de Ciencias Básicas, Facultad de
      Ingeniería, Universidad Nacional de La Plata, C.C. 67, 1900 La
      Plata, Argentina}
\author{C. Iemmi}
    \affiliation{Departamento de Física, Facultad de Ciencias Exactas
      y Naturales, Universidad de Buenos Aires, 1428 Bueno Aires, Argentina}
    \affiliation{Consejo Nacional de Investigaciones Cient\'ificas y
      T\'ecnicas, 1425 Buenos Aires, Argentina}


\begin{abstract}
  The temporal evolution of a quantum system can be characterized by
  quantum process tomography, a complex task that consumes a number of
  physical resources scaling exponentially with the number of
  subsystems.  An alternative approach to the full reconstruction of a
  quantum channel allows selecting which coefficient from its matrix
  description to measure, and how accurately, reducing the amount of
  resources to be polynomial. The possibility of implementing this
  method is closely related to the possibility of building a complete
  set of mutually unbiased bases (MUBs) whose existence is known only
  when the dimension of the Hilbert space is the power of a prime
  number. However, an extension of the method that uses tensor
  products of maximal sets of MUBs, has been introduced recently.
  Here we explicitly describe how to implement this algorithm to
  selectively and efficiently estimate any parameter characterizing a
  quantum process in a non-prime power dimension, and we conducted for
  the first time an experimental verification of the method in a
  Hilbert space of dimension $d=6$. That is the small space for which
  there is no known a complete set of MUBs but it can be decomposed as
  a tensor product of two other Hilbert spaces of dimensions $D_1=2$
  and $D_2=3$, for which a complete set of MUBs is known. The
  $6$-dimensional states were codified in the discretized transverse
  momentum of the photon wavefront. The state preparation and
  detection stages are dynamically programmed with the use of
  only-phase spatial light modulators, in a versatile experimental
  setup that allows to implement the algorithm in any finite
  dimension.
\end{abstract}


\maketitle

\input{sections/intro.tex}
\input{sections/method_description.tex}

\input{sections/experimental_implementation.tex}
\input{sections/results_and_discussion.tex}

\input{sections/conclusions.tex}

\begin{acknowledgments}
This work was supported by Universidad de Buenos Aires
(UBACyT Grant No.~20020170100564BA). Q.P.S. was supported
by a CONICET Fellowship.
\end{acknowledgments}
\appendix*
\input{sections/appendix.tex}
\bibliography{references}

\end{document}

%% file: preamble.tex
\usepackage{amsthm}
\usepackage{mathtools}
\usepackage{physics}
\usepackage{xcolor}
\usepackage{graphicx}
\usepackage[left=23mm,right=13mm,top=35mm,columnsep=15pt]{geometry} 
\usepackage{adjustbox}
\usepackage{placeins}
\usepackage[T1]{fontenc}
\usepackage{lipsum}
\usepackage{csquotes}

%% file: auxcommands.tex
\renewcommand{\ket}[1]{\ensuremath{|#1\rangle}}
\renewcommand{\bra}[1]{\ensuremath{\langle #1|}}

\renewcommand{\ketbra}[2]{\ensuremath{|#1\rangle\!\langle#2|}}

\newcommand{\Proj}[1]{\ensuremath{\ketbra{#1}{#1}}}

\newcommand{\fid}{\bar F}
\newcommand{\E}{\mathcal{E}}

\newcommand{\Id}{\mathbb{I}}


%% file: sections/intro.tex
\section{Introduction}\label{sec:intro}
The research in the field of quantum information processing is
continuously growing, mainly driven by promising technological
applications, that range from quantum computation, to quantum
cryptography and communication
\cite{Outeiral2020,GoogleAIQuantum2020,Liao2017,Gisin2007,Llewellyn2020}.
On the way to developing reliable quantum technologies, it becomes
crucial the ability to characterize an unknown quantum device, a task
commonly referred as quantum process tomography (QPT)
\cite{Mohseni2008}. This technique is specially useful to
experimentally characterize the decoherence mechanism that take place
in noisy quantum gates~\cite{Kofman2009}. For instance, once a given
quantum device has been characterized, the \textit{a priori} knowledge
of the temporal evolution of any quantum state could be used to design
error correction schemes \cite{Devitt2013}.

In this context, different QPT schemes have been tested experimentally
for diverse physical implementations of quantum systems: polarization
of photons \cite{Altepeter2003, Kim2018, Wang2007}, superconducting
qubits \cite{Bialczak2010,Yamamoto2010}, nuclear magnetic-resonance
quantum computers \cite{Childs2001}, and ion traps \cite{Riebe2006},
among others. However, since this is considered a \textit{hard task}
due to the required physical resources, the research for efficient
schemes becomes more and more relevant as the size of experimentally
feasible systems increases.

Within the formalism of quantum mechanics the state of a physical
system is described by a density matrix $\rho$, and the quantum
operation of a device can be mathematically represented by a linear,
completely positive map, $\E$, that applied over a quantum state
$\rho_{in}$ returns the state
$\rho_{out}= \E \left(\rho_{in}\right)$~\cite{Nielsen}. The effect of
this map can always be written in the \textit {so-called operator-sum
  representation} or \textit {Kraus decomposition} as
 \begin{equation} \mathcal{E} \left(\rho\right)=\sum_{i}A_i \rho
A_i^\dagger,
\label{eq:Mapa-kraus}
\end{equation} where $\{A_i\}_i$ is a set of linear operators that act
on a Hilbert space $\mathcal{H}$, and 
satisfy the relation $\sum_{i}A_i A_i^\dagger\leq\Id$. If the
dimension of the system under consideration is $d$, one can choose a basis
of operators, $\left\lbrace E_m, m=0,\ldots,d^2-1\right\rbrace$, and rewrite Eq.~(\ref{eq:Mapa-kraus}) as
\begin{equation} \mathcal{E} \left(\rho\right)=\sum_{mn}\chi_{mn}E_m
\rho E_n^\dagger,
\label{eq:Mapa}
\end{equation}
where $\chi$ is an Hermitian and positive matrix and the trace
preserving condition is given by
$\sum_{mn}\chi_{mn}E_n^\dagger E_m=\Id$.
Once the operator basis $\left\lbrace E_m\right\rbrace$ is fixed,
performing QPT is equivalent to determining the matrix coefficients
$\chi_{mn}~'s$. Therefore, the full characterization of the map
requires $d^4-d^2$ real parameters, and in the case of $n$-qubit
systems, this is associated with an exponentially large number of
coefficients to be determined ($d=2^n$). Moreover, standard methods
require an amount of experimental and computational resources that
scale exponentially with the number $n$ of subsystems, even to
determine a single coefficient.
In this context, a protocol for quantum process tomography is said to
be:
\begin{itemize}
\item\textit{selective}, if it allows to obtain, individually, the
  coefficients of the matrix $\chi$, i.e, without having to perform
  the full QPT in case we are only interested in some particular
  element $\chi_{mn}$, and
\item\textit{efficient}, if any coefficient $\chi_{mn}$ can be determined with sub-exponential resources.  
\end{itemize}

In previous works~\cite{Bendersky2008,Bendersky2009}, a protocol for
selective and efficient quantum process tomography (SEQPT) was
developed and successfully accomplished experimentally on different
physical platforms~\cite{Schmiegelow2011,Gaikwad2018}. However, the
protocol is implementable as long as the dimension of the Hilbert
space is the power of a prime number. This is because the SEQPT makes
use of a complete set of mutually unbiased bases (MUBs), a
construction which is only known to exist for prime-power dimensions
\cite{Wootters1989, Ivonovic1981, Durt2010}.

More recently, two schemes that allow extending the SEQPT protocol to
arbitrary finite dimensions were presented in
Ref.~\cite{Perito2018}. One of these scheme is based on tensor
products of complete sets of MUBs in lower prime-power dimensions, as
a good approximation to solve the original problem. The other one
starts from a complete set of MUBs in a higher dimension, and then
projects this set onto the desired dimension. Which strategy to follow
will depend mainly on the physical implementation: the SEQPT with
tensor product, for example, requires the preparation of product
states in smaller dimensions and it could be the most suitable option
for composite systems, although there is no advantage over the SEQPT
with projection in relation to the number of individual experiments
required to estimate a given coefficient.

In this work, we present for the first time the experimental
realization of the tensor product scheme for the SEQPT protocol.  The
method is applied to characterize a trace preserving quantum process
on dimension $d=6$, that is, the smallest Hilbert space for which the
protocol for SEQPT in non-power prime dimension becomes relevant.
Among the many possible codifications for a quantum state of dimension
$d$ (qudit), the spatial degrees of freedom of a single photon provide
an easy access to dimensions $d\geq 2$. In particular, here we have
encoded the $d$-dimensional system in the discretized transverse
momentum of single photons, a scheme widely used for implementing
quantum information processing in high-dimension
\cite{Canas2014b,Etcheverry2013,Solis-Prosser2017} and which has
proven useful for testing protocols for both, quantum state
tomography~\cite{Goyeneche2015,PearsStefano2019} and quantum process
tomography~\cite{Varga2018}.

The paper is organized as follows: in Section \ref{sec:method} we
briefly describe the main idea behind the SEQPT protocols, with
particular emphasis in the tensor product scheme for the case of
dimensions with two different prime numbers in its factorization,
given that this will be the case in which we will focus through our
experiment. After that, in Section \ref{sec:experiment}, we describe
the experimental setup and, finally, in Section
\ref{sec:resultsanddiscussion} we present our results and conclusions.

%% file: sections/method_description.tex
\section{SEQPT method}\label{sec:method}

Let us first briefly review the theoretical background for the SEQPT
protocol in prime power dimensions \cite{Bendersky2008,Bendersky2009}
and its generalization to a more general case when the dimension is
factorized as a product of two prime power
dimensions~\cite{Perito2018}.

\subsection{Haar integrals of quadratic forms and 2--designs}

The protocol for SEQPT that we will implement in this work is based on
the following properties:
\begin{itemize}
\item For any two operators $A$ and $B$ in a Hilbert space
  $\mathcal H$ of dimension $d$, it holds that
\begin{equation}\label{eq:traces}
  \int_\mathcal H d\psi\, \Tr[P_\psi\,A\,P_{\psi}B]=\frac{\Tr A\; \Tr B+\Tr[AB]}{d(d+1)} \, ,
\end{equation}
where $P_\psi=\ketbra{\psi}{\psi}$, and the integration is performed
over the only normalized unitarily invariant measure on $\mathcal H$,
namely, the Haar measure.

\item A finite set of states
  $\mathcal X=\left\lbrace\ket{\psi_m}, m=1,...,N\right\rbrace$ is a
  uniform state 2--design if:
\begin{equation}
 \int_\mathcal H d\psi\,f\left(P_\psi\right)= \frac{1}{N}\sum_{m=1}^N  f\left(P_{\psi_m}\right)\label{2designInt}
\end{equation}
for any $f$ that is quadratic in $P_\psi$, and the integration
is performed  again over the Haar measure.
\end{itemize}%
Therefore, a state 2--design is a set of states on which the mean
value of any quadratic function in $P_\psi$ gives the same mean value
as on the set of all possible states in $\mathcal{H}$.  Note that, in
particular, this kind of sets allows to easily compute quantities a in
Eq.~(\ref{eq:traces}).

\subsection{Quantum channel fidelity and SEQPT in prime power dimension}
\label{sec:seqpt}

Given a quantum channel $\mathcal E$, its mean fidelity is given by:
\begin{equation}
\bar F(\mathcal E)=\int_\mathcal H d\psi \; \Tr[P_\psi \, \E \left(P_{\psi}\right)] \, ,
\end{equation}
where the integration is taken over the Haar measure.
According to Eq.~\eqref{eq:Mapa} we will expand $\mathcal E$ by
selecting an operator basis $\{E_m\}$ that is orthogonal
($\Tr(E_mE_n^\dagger)=d\,\delta_{m,n}$) and unitary
($E_n E_n^\dagger = \mathbb{I}$). If we define the modified channel as
\begin{eqnarray}
  \mathcal E_{ij}(\rho)\equiv\E(E^\dagger_i \rho E_j), 
\end{eqnarray}
a direct application of the property given by Eq.~\eqref{eq:traces}
relates the mean fidelity of $\mathcal E_{ij}$ with the element
$\chi_{ij}$ of the matrix description of the channel $\mathcal E$. We
will focus here in trace preserving maps, i.e., where the condition
$\sum_iA_iA_i^{\dagger}=\Id$ is hold. In such particular case, the
relation is explicitly
\begin{equation}
\label{eq:meanfid}
\bar F(\E_{ij})= \frac{d\,\chi_{ij}+\delta_{ij}}{d+1} \, .
\end{equation}
Moreover, given that the mean fidelity is the integral over the Haar
measure of a quadratic form in $P_\psi$, it can be computed just by
evaluating and averaging the survival probability, through the channel
$\mathcal E_{ij}$, over the states of a 2--design
$\bar
F(\E_{ij})=\frac{1}{N}\sum_{m=1}^N\Tr\left[\ket{\psi_m}\bra{\psi_m}~\E\left(E^\dagger_i
    \ket{\psi_m}\bra{\psi_m} E_j\right)\right]$.  This finally gives
the clue along with Eq.~(\ref{eq:meanfid}) to design the experiments
to find the desired coefficients $\chi_{ij}$, once a a 2--design is
known.

A simple way to find a state 2--design is to consider a set of $(d+1)$
MUBs, which automatically form a state 2--design
\cite{Klappenecker2005} and their construction is known when the
dimension $d$ is the power of a prime number
\cite{Ivonovic1981,Wootters1989}. However, for arbitrary dimension
$d$, it is not known the maximum number of MUBs.

\subsection{SEQPT in arbitrary finite dimension}

In the general case, the previous protocol fails because of the lack
of a uniform 2--design when the dimension of the system $d$ is not the
power of a prime number. However, two strategies that allow the
generalization of the SEQPT protocol to an arbitrary dimension were
recently presented in Ref.~\cite{Perito2018}. They consist in finding
a finite set of states that, despite not being a uniform 2--design,
allows to compute mean fidelities in a reasonable way. In particular,
we will follow the \text{tensor product approach} based on the fact
that tensor products of 2-designs can be used to approximate
2-designs. Since an arbitrary dimension $d$ can always be factorized
into power of prime numbers, then the tensor products of maximal MUB
sets provide a good approximation for integration purposes.

We will focus on the bipartite case in which we are concerned in this
work. In such a case, the dimension of the Hilbert space $d$ is
factorized as $d=D_1 D_2$ where $D_1=p_1^{n_1}$, $D_2=p_2^{n_2}$, and
$p_1$, $p_2$ are prime numbers. The first step is to expand the
channel $\mathcal E$ in a basis that is a product of operators acting
on $\mathcal{H} = \mathcal{H}_1 \otimes \mathcal{H}_2$, where the
dimensions of the subsystems are $D_1$ and $D_2$, respectively. This
basis can be written in terms of two orthogonal operator bases
$\{ E_{j_1j_2} \equiv E_{j_1} \otimes E_{j_2}
\}_{j_1=0,\dots,D_1^2-1}^{j_2=0,\dots,D_2^2-1}$, where each element
$E_{j_i}$ ($i=1,2$) is an unitary matrix.  Thus, the expansion in
Eq.~(\ref{eq:Mapa}) is rewritten as
\begin{equation}
  \mathcal E (\rho) =  
  \sum_{\mu_1\mu_2\nu_1\nu_2} \chi_{\nu_1\nu_2}^{\mu_1\mu_2} E_{\mu_1\mu_2} \rho   E^{\nu_1\nu_2}   
  \, ,
\label{canal2primos}
\end{equation}
for some coefficients $\chi_{\nu_1\nu_2}^{\mu_1\mu_2}$. We have
adopted the convention $E^{j_i}\equiv E_{j_i}^\dagger$,
$E^{j_1j_2} \equiv E_{j_1j_2}^\dagger$, and hereafter, we will also
consider $\delta_i^j\equiv\delta_{ij}$. If we take $X_1$ and $X_2$ as
2--designs in $\mathcal{H}_1$ and $\mathcal{H}_2$, respectively, and
define $X_\otimes \subset \mathcal{H}_1 \otimes \mathcal{H}_2$ as the
set of all possible tensor product between states in $X_1$ and states
in $X_2$, the coefficient $\chi_{j_1j_2}^{i_1i_2}$ can be expressed,
as
\begin{eqnarray}
\label{eq:chifid}
\chi_{j_1j_2}^{i_1i_2}&=&\fid_\otimes(\E_{j_1j_2}^{i_1i_2})\frac{(1+D_1)(1+D_2)}{d}+\frac{\delta^{
i_1}_{j_1}\delta^{i_2}_{j_2}}{d}\\
&-&\fid_1(\mathcal E_{j_1j_2}^{i_1i_2})\frac{(1+D_1)}{d}-\fid_2(\E_{j_1j_2}^{i_1i_2})\frac{(1+D_2)}{d},\nonumber
\end{eqnarray}
where the modified channel is now given by
$\E_{j_1j_2}^{i_1i_2}(\rho) = \E(E^{i_1i_2} \rho E_{j_1j_2})$. The mean fidelity of this modified channel 
is expressed as
\begin{multline}
\bar F_\otimes(\E_{j_1j_2}^{i_1i_2})=\\
\int_{\mathcal H_1} \int_{\mathcal H_2} d\psi_1 d\psi_2\,\Tr[P_{\psi_1\psi_2} \,\E_{j_1j_2}^{i_1i_2} \left(P_{\psi_1\psi_2}\right)],
\end{multline}
%
and this double integral can be evaluated by averaging over
the finite set of states in $X_\otimes$:
\begin{eqnarray}
&&\bar F_\otimes(\E_{j_1j_2}^{i_1i_2})=
\nonumber\\
&&\frac{1}{\left | X_1 \right |}\frac{1}{\left | X_2 \right |}
\sum_{\ket{\psi_1}\in X_1}\sum_{\ket{\psi_2}\in X_2}\Tr[P_{\psi_1\psi_2} \,\E_{j_1j_2}^{i_1i_2} \left(P_{\psi_1\psi_2}\right)]\nonumber
\\
&~&=\frac{1}{\left | X_\otimes \right |}
\sum_{\ket{\psi}\in X}\Tr[P_{\psi} \,\E_{j_1j_2}^{i_1i_2} \left(P_{\psi}\right)].\label{mean_fidelity}
\end{eqnarray}
\begin{figure}
\begin{center}
\includegraphics[width=.4\textwidth]{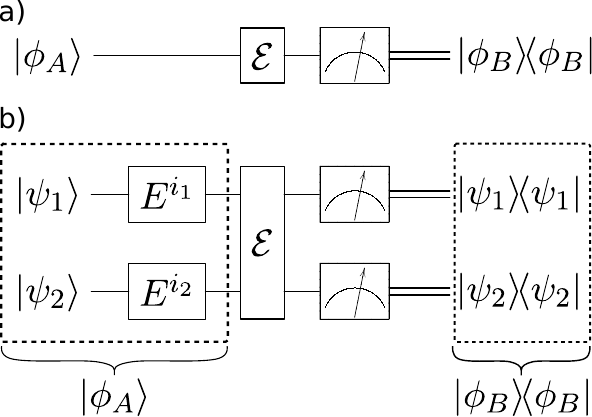}
\caption{\textbf{a)} Circuit for a projective measurement of state
  $\ket{\phi^A}$, after being affected by the process $\E$, onto the
  state $\ket{\phi^B}$. \textbf{b)} Circuit for measuring the survival
  probability of the state
  $\ket{\psi}=\ket{\psi_1}\otimes\ket{\psi_2}$ through the modified
  channel $\E_{i_1i_2}^{i_1i_2}$. By sampling over $X_{\otimes}$ we
  can obtain the diagonal element $\chi_{j_1j_2}^{i_1i_2}$
  corresponding to the process matrix of $\E$.\label{fig:circuit}}
\end{center}
\end{figure}%
Furthermore, measuring the action of the modified channel over the
states in $X_\otimes$ is enough to compute, not only
$\bar F_\otimes(\E_{j_1j_2}^{i_1i_2})$, but {\it all} the terms in
Eq.~\eqref{eq:chifid}. For instance, in order to estimate the reduced
mean fidelity over subsystem $X_1$
\begin{multline}
\fid_1(\E_{j_1j_2}^{i_1i_2})=\\\int_{\mathcal{H}_1} d\psi_1 \bra{\psi_1}\Tr_2\left[\E_{j_1j_2}^{i_1i_2}(P_{\psi_1}\otimes\Id_2/D_2)\right]\ket{\psi_1} \, ,
\end{multline}
one has to measure the survival expectation of
$ P_{\psi_1} \otimes \Id_2$ given the initial state
$ P_{\psi_1} \otimes \frac{\Id_2}{D_2}$, that is
%
\begin{multline}\label{eq.Fid1}
\fid_1(\E_{j_1j_2}^{i_1i_2})=\\\frac{1}{\left | X_1 \right |}
\sum_{\ket{\psi_1}\in X_1}\Tr\left[\left(P_{\psi_1}\otimes \Id_2\right)\E_{j_1j_2}^{i_1i_2}\left(P_{\psi_1}\otimes \Id_2/D_2\right)\right].
\end{multline}
%
It can be achieved by looking at the statistics of the measurements on
system 1 independently from the results of the measurements on system
2, and similarly for $\fid_2(\E_{j_1j_2}^{i_1i_2})$. This is because
the initial state of system 2 is a random state from $(1+D_2)$
orthogonal bases, which is a possible implementation of $\Id_2$
provided the result of the measurement of system 2 is not taken into
account.
Thus, the selectivity of the method is given by the fact that a
particular element $\chi_{i_1i_2}^{j_1j_2}$ can be determined by
calculating the three mean fidelities $\fid_\otimes$, $\fid_1$ and
$\fid_2$, over the modified channel
$\E_{i_1i_2}^{j_1j_2}$. Furthermore, this fidelities can be estimated
\emph{efficiently} by randomly sampling states in $X_\otimes$: given a
fixed error tolerance, the number of states to be sampled is
independent of the dimension.

Figure~\ref{fig:circuit} depicts the procedure to follow in the
reconstruction of a given coefficient $\chi_{j_1j_2}^{i_1i_2}$. Let us
assume that we have an experimental setup described by the circuit in
Fig.~\ref{fig:circuit} a) where an arbitrary state $\ket{\phi^A}$ is
prepared and, after being affected by the process $\mathcal{E}$, it is
projected onto the state $\ket{\phi^B}$.
\\\\
\textbullet\ \ \emph{Diagonal case:}\ %
for $i_1=j_1$ and $i_2=j_2$, the effect of the modified channel
on the state
$\ket{\psi}=\ket{\psi_1}\otimes\ket{\psi_2}\in X_{\otimes}$, is
\begin{eqnarray}
\E_{i_1i_2}^{i_1i_2}(\ketbra{\psi}{\psi}) &=& \E(E^{i_1i_2} \ketbra{\psi}{\psi}E_{i_1i_2})\nonumber\\
&=&\E(E^{i_1}P_{\psi_1}E_{i_1} \otimes E^{i_2}P_{\psi_2}E_{i_2} ), 
\end{eqnarray}
and the survival expectation can be obtained by performing a
projective measurement onto $\ket{\psi_1}\otimes\ket{\psi_2}$. The
circuit describing this procedure is shown in Fig.~\ref{fig:circuit}
b), where now the input state is $\ket{\phi^A} = E^{i_1i_2}\ket{\psi}$
and the state to be projected onto is $\ket{\phi^B} = \ket{\psi}$. An
explicit construction of the 2--designs $X_1$, $X_2$ and the
corresponding operator bases is discussed in
Appendix~\ref{sec:appendix}.
\\\\
\textbullet\ \emph{Non-diagonal case:\ }%
 for $i_1\neq j_1$ or $i_2\neq j_2$, the resulting modified channel is non-physical. In fact, its effect on the sampled state $\ket{\psi}$ is given by
\begin{eqnarray}
\E_{j_1j_2}^{i_1i_2}(\ketbra{\psi}{\psi}) &=& \E(E^{i_1i_2} \ketbra{\psi}{\psi}E_{j_1j_2})\nonumber\\
&=&\mathcal{E}(\ketbra{\alpha}{\beta}),
\end{eqnarray}
whith $\ket{\alpha} = E^{i_1i_2}\ket{\psi}$ and
$\ket{\beta} = E_{j_1j_2}\ket{\psi}$. This is equivalent to the action
of the original channel $\mathcal{E}$ on the matrix
$\ketbra{\alpha}{\beta}$, which is not a density matrix, and therefore
does not represent a physical state. However, this matrix can always
be expressed as a linear combination of \textit{at most} five matrices, each corresponding to a projector. If $\ket{\alpha}$ and $\ket{\beta}$ are
orthonormal,
$\mathcal{E}(\ketbra{\alpha}{\beta}) = \mathcal{E}(\Proj{+}) +
\mathcal{E}(\Proj{-}) - \frac{1+i}{2} \left(\mathcal{E}(\Proj{\alpha})
  + \mathcal{E}(\Proj{\beta})\right),$ with
$\ket{+} = (\ket{\alpha} + \ket{\beta})/\sqrt{2}$ and
$\ket{-} = (\ket{\alpha} + i \ket{\beta})/\sqrt{2}.$ If they are not
orthonormal, a similar decomposition exists.  Then, the linearity of
$\E$ ensures that we can compute the action of the modified channel $\E_{j_1j_2}^{i_1i_2}$
over any state as a linear combination of the action of the original channel $\mathcal{E}$
over a suitable choice of pure states.

%% file: sections/experimental_implementation.tex
\section{Experimental tensor product SEQPT} \label{sec:experiment}

In order to experimentally test the tensor product \mbox{SEQPT}
protocol, we have implemented and reconstructed a quantum process
$\mathcal{E}$ in a Hilbert space of dimension $d=6$, for which a
maximal sets of MUBs is not known. In our case, $\mathcal{E}$ is a non
trivial process over qudit states encoded in the discretized
transverse position of single photons \cite{Neves2005}.

\begin{figure}[ht]
\begin{center}
\includegraphics[width=.5\textwidth]{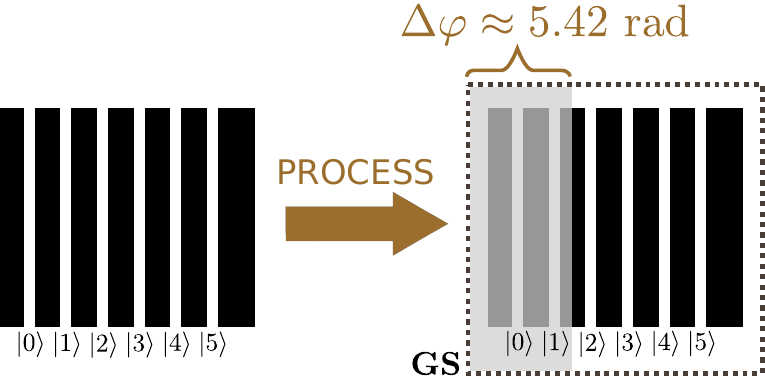}
\caption{Effect of the process on the $6$-dimensional spatial qudit
  codified in the discretized transverse momentum of a photon. The
  process is physically implemented by a glass slab (GS) with a
  transparent coating covering part of its surface. This partial
  coating introduces a phase shift $\Delta\varphi = 5.42~\mathrm{rad}$
  on the paths that codify the states $\ket{0}$ and $\ket{1}$, of the
  $6$-dimensional canonical basis. \label{fig:process-implementation}}
\end{center}
\end{figure}

For this encoding, a $d$-dimensional quantum state can be defined by
means of a complex aperture consisting of $d$ slits and placed in the
path propagation of the photon field, so that, the dimension of the
spatial qudit is determined by the number of paths available to the
photon. To be more specific, when such an aperture is illuminated by a
paraxial and monochromatic single photon field, which is approximately
constant on the aperture area, the resulting state--usually called
\emph{slit state}-- can be described by
\begin{eqnarray}
\ket{\psi} =\frac{1}{\mathcal{N}} \sum_{k=0}^{d-1} c_k \ket{k}, 
\label{6state}\end{eqnarray}
where $c_k$ is the complex transmission of the $k$-th slit, $\ket{k}$
represents the transverse-path state of a single photon trough this
slit, and the normalization constant is given by
$\mathcal{N}=\sqrt{\sum_{i=0}^{d-1}|c_i|^2}$.  Optically, $|c_k|^2$
and $\arg(c_k)$ correspond to the intensity transmission and phase
retardation of the $k$-slit, which can be controlled, independently,
defining the complex aperture by means of programmable spatial light
modulators (SLMs)~\cite{Lima2009,Solis-Prosser2013}.

We made the following assignments between states in the canonical basis of
$d=6$ to the tensor product of elements of the canonical basis
of $D_1=2$ and $D_2=3$:
\begin{eqnarray}
\begin{matrix}
\ket{0}\rightarrow \ket{0}\otimes\ket{0}\;\;, & \ket{3}\rightarrow \ket{1}\otimes\ket{0}\\   
\ket{1}\rightarrow \ket{0}\otimes\ket{1}\;\;, & \ket{4}\rightarrow \ket{1}\otimes\ket{1} \\ 
\ket{2} \rightarrow \ket{0}\otimes\ket{2}\;\;, & \ket{5}\rightarrow \ket{1}\otimes\ket{2} 
\end{matrix}\label{2x3state}
\end{eqnarray}
and according to this, the state in Eq.~(\ref{6state}) is rewritten as
\begin{eqnarray}
\ket{\psi} =\frac{1}{\mathcal{N}} \sum_{k_1=0}^{D_1}\sum_{k_2=0}^{D_2} c_{k_1k_2} \ket{k_1}\otimes\ket{k_2}. 
\label{6state_b}\end{eqnarray}

The target process to be implemented corresponds to adding a constant phase
shift, $\Delta\varphi$, to the states $\ket{0}$ and $\ket{1}$ of the canonical basis in $d=6$. For this process, a decomposition in terms
of Kraus operators is 
\begin{equation}
\begin{split}
\mathcal{E}_t(\rho)&=A_t\rho A_t^{\dagger},\\
A_t& = e^{i\Delta\varphi}\left(|0\rangle\langle0| + |1\rangle\langle1| \right) +\sum_{k=2}^{5}|k\rangle\langle k|.
\end{split}\label{Kraus-decomp}
\end{equation}
Physically, this was realized by means of a rectangular glass slab (GS) partially coated with a
transparent material, resulting in an extra phase
of $\Delta\varphi = 5.42\ \mathrm{rad}$ for the wavelength used in our
experiment. Figure \ref{fig:process-implementation} shows, schematically,
the effect of the target process $\mathcal{E}_t$ when acting on a state generated by 6-slit aperture.

\begin{figure}
\includegraphics[width=.5\textwidth]{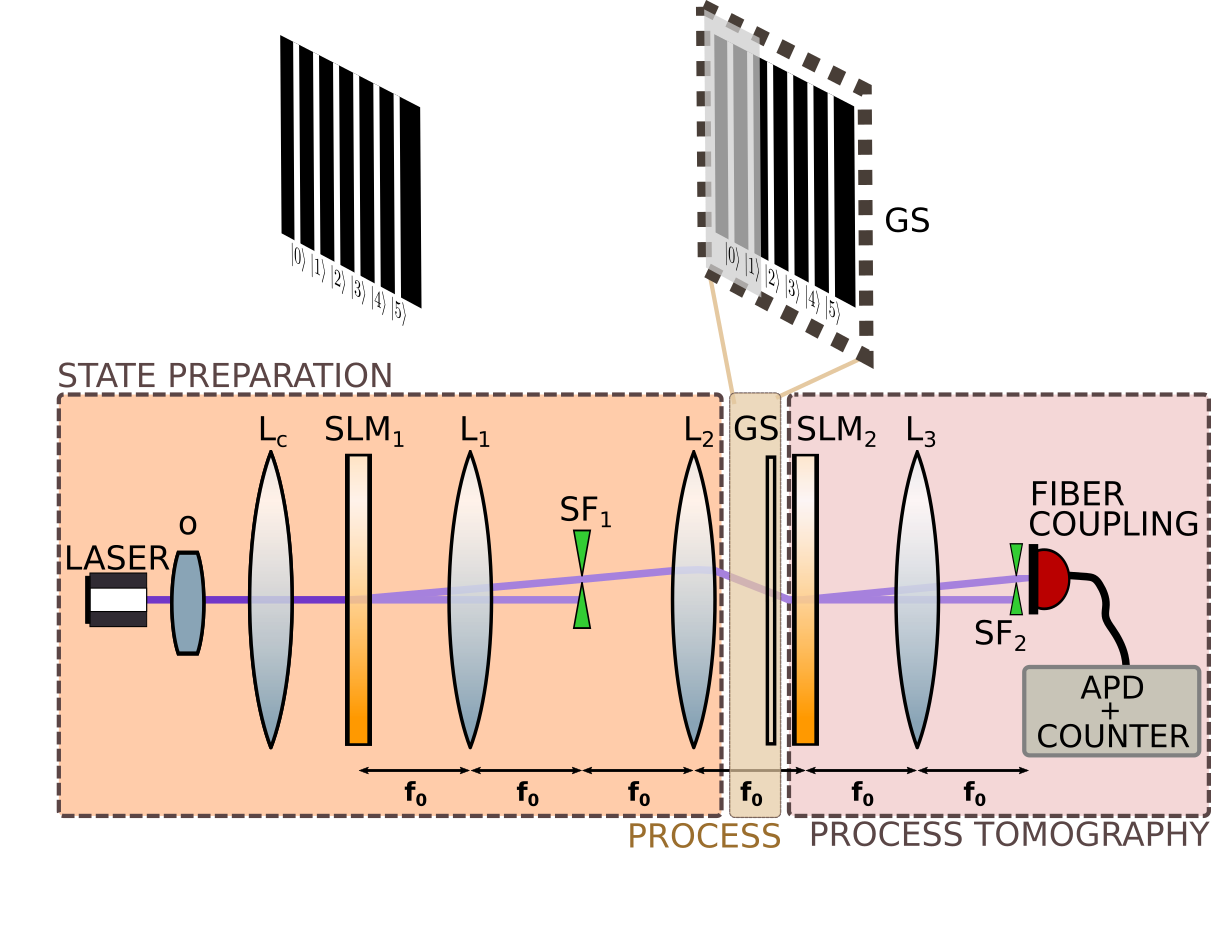}
\caption{Experimental setup. A 405nm cw laser diode is attenuated to the single
photon level. O: microscope objective; Ls: convergent lenses; SLMs: pure phase spatial light
modulators; SFs: spatial filters. A glass slab (GS) implements the process on a spatial qudit as schematized in  Fig.~\ref{fig:process-implementation}. The detection in the
centre of the interference pattern is performed with a
fiber-coupled APD.}\label{fig:experimental-setup}
\end{figure}

The experimental setup is shown in
Fig.~\ref{fig:experimental-setup}. It is a flexible configuration that
allows us to generate arbitrary pure states and perform general
projective measurements based in the use of phase-only
SLMs~\cite{Solis-Prosser2013}.  It can be divided in two main parts:
the state preparation (SP) part, in which the state $\ket{\phi^A}$,
that subsequently cross the channel, is prepared, and the process
tomography part, where is selected the state $\ket{\phi^B}$ onto which
$\mathcal{E}_t(\ket{\phi^A}\bra{\phi^A})$ is finally projected.

Let us describe the SP part.  The light source is a laser diode
$@405\mathrm{nm}$, that is expanded and collimated by the microscope
objective O and the lens $\mathrm{L_c}$, respectively. A neutral
density filter (not shown in Fig.~\ref{fig:experimental-setup})
attenuates the laser down to the single photon level. The complex
aperture that generates each qudit $\ket{\phi^A}$ is displayed in the
phase-only $\mathrm{SLM_1}$, that is uniformly illuminated by the
collimated incoming beam. This SLM consists in a twisted nematic
liquid crystal display (LCD) Sony LCX012B coupled to polarizers and
wave plates. By selecting suitable polarization
states~\cite{Marquez2001}, both at the input and the output of the
LCD, a \emph{phase-only modulation} of $2\pi@405\mathrm{nm}$ on the
wavefront, is attains. This LCDs have a VGA resolution
($640\times480$) with pixels of $43\mathrm{\mu m}$. The displayed
slits were defined to have a width of $4\ \mathrm{pixels}$ and a
separation of 6 $\mathrm{pixels}$ between their centers.

To control independently the complex amplitude of every slit
--transmisivity and phase retardation-- with a phase-only SLM, we
implement the method described in the
Ref.~\cite{Solis-Prosser2013}. Briefly, this is achieved by
programming a different-phase grating in the spatial region
corresponding to each slit. The depth of the grating determines the
efficiency in the first diffraction order, which codified the real
amplitude $|c_k|$ of the superposition in Eq.~(\ref{6state}), while a
constant phase added per slit defines its complex argument
$\mathrm{arg}(c_k)$. The lenses $\mathrm{L_1}$ and $\mathrm{L_2}$
(both of focal length $f_0=26 \mathrm{cm}$), together with the spatial
filter $\mathrm{SF_1}$, form the $4-f$ optical processor that select
this diffracted order. Thus, at the back focal plane of $\mathrm{L_2}$
the wavefront distribution corresponds to the the desired spatial
qudit.

\begin{figure*}[ht!]
\includegraphics[width=\linewidth]{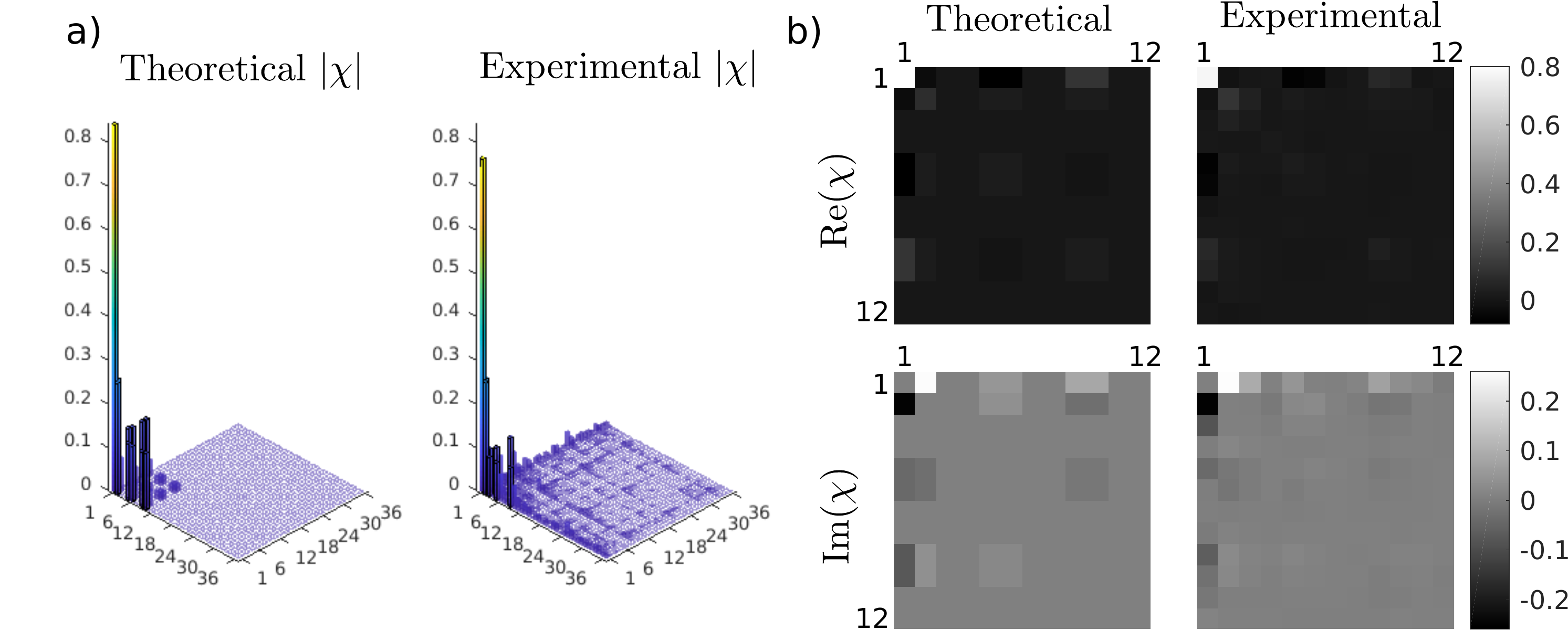}
\centering 
\caption{\textbf{a)} Comparison of the absolutes values of the elements
  of $\chi_\mathrm{theo}$ (expected matrix of the target process $\mathcal{E}_t$) and $\chi_\mathrm{exp}$ (reconstructed by
  the tensor product SEQPT). \textbf{b)} Detail of the absolute value of the elements in the $12\times12$ sub-matrix. This elements correspond to the non-zero block in the matrix $\chi_\mathrm{theo}$ (upper-left block in the theoretical plot of \textbf{a)}). The gray-scale map shows the real and imaginary part for the
  theoretical and the experimental reconstructed
  matrices.\label{fig:rho}}
\end{figure*}

In the PT part, a second SLM ($\mathrm{SLM_2}$) with similar
characteristics to those of $\mathrm{SLM_1}$ and operating in the same
way, encodes each projection base state $\ket{\phi^B}$. In the absence
of channel $\mathcal{E}_t$ carried on by means of GS, if
$\ket{\phi^B} = \sum b_k \ket{k}$ the resulting state after
$\mathrm{SLM_2}$ is proportional to $\sum c_k b_k^{*} \ket{k}$. This
second SLM is placed at the front focal plane of lens
$\mathrm{L_3}$. After filtering the first diffracted order by means of
$\mathrm{SF_2}$, the exact Fourier transform of the projected spatial
qudit is obtained at the detector plane. The light distribution
corresponds to the interference pattern projection between the
prepared state and the selected projector state. The light of the
center of this pattern is coupled by a single-mode fiber into a single
photon counting module Perkin Elmer SPCM-AQRH-13-FC, based on an
avalanche photodiode (APD). Then, the single photon count rate is
proportional to the probability of projection of the two states,
$p(\ket{\phi^A}, \ket{\phi^B})$ \cite{Lima2011}. In the presence of
channel $\mathcal{E}_t$, this probability is now
$p(\E(\ketbra{\phi^A}{\phi^A}), \ket{\phi^B})$.

%% file: sections/results_and_discussion.tex
\section{Results and Discussion} \label{sec:resultsanddiscussion}

To evaluate the viability of the method, we first performed the full
tomography of the process $\mathcal{E}_t$ introduced in our
experimental setup by means of GS. To this end we have reconstructed
each element of the matrix, $\chi_{j_1j_2}^{i_1i_2}$, by averaging
over all the elements of the tensor product of the 2-design
$X_{\otimes}$ (see Eqs.~(\ref{eq:chifid})-(\ref{eq.Fid1})). The
experimental matrix $\chi_\mathrm{exp}$ was post-processed with the
complete positive trace preserving projection (CPTP) algorithm
presented in Ref.~\cite{Knee2018}. This projection ensures that the
resulting matrix is completely positive. Then, it represents a
physical process and the trace of any quantum state is preserved.  The
last constraint is in agreement with the target process
$\mathcal{E}_t$. Figure~\ref{fig:rho}.a) shows the comparison between
the absolute values of the elements of theoretical matrix
$\chi_\mathrm{theo}$ and $\chi_\mathrm{exp}$ in the measurement
basis. For a better comparison, Fig.~\ref{fig:rho}.b) shows a detail
of the non-zero $12\times12$ block of the expected matrix comparing
both the real and imaginary part of $\chi_\mathrm{theo}$ and
$\chi_\mathrm{exp}$. As figure of merit and resorting to the
Choi-Jamiolkowski isomorphism \cite{Mohseni2008}, we calculate the
similitude between $\chi_\mathrm{theo}$ and $\chi_\mathrm{exp}$ as the
fidelity
$F\equiv
F(\rho_\mathrm{theo},\rho_\mathrm{exp})=\mathrm{Tr}\sqrt{\sqrt{\rho_\mathrm{theo}}\rho_\mathrm{exp}\sqrt{\rho_\mathrm{theo}}}$
between two quantum states, $\rho_\mathrm{theo}$ and
$\rho_\mathrm{exp}$, assigned to the target process and to the
experimentally reconstructed one $(\mathcal{E}_{\mathrm{exp}})$,
respectively. The obtained value is $F \approx 0.93$. For completeness
and to make the reconstruction quality of the method independent of
the errors inherent to the experimental setup, we have also performed
a standard QPT~\cite{Nielsen}, and as a result, a comparable fidelity
value for this reconstruction method was obtained.

\begin{figure}[t]
\includegraphics[width=\linewidth]{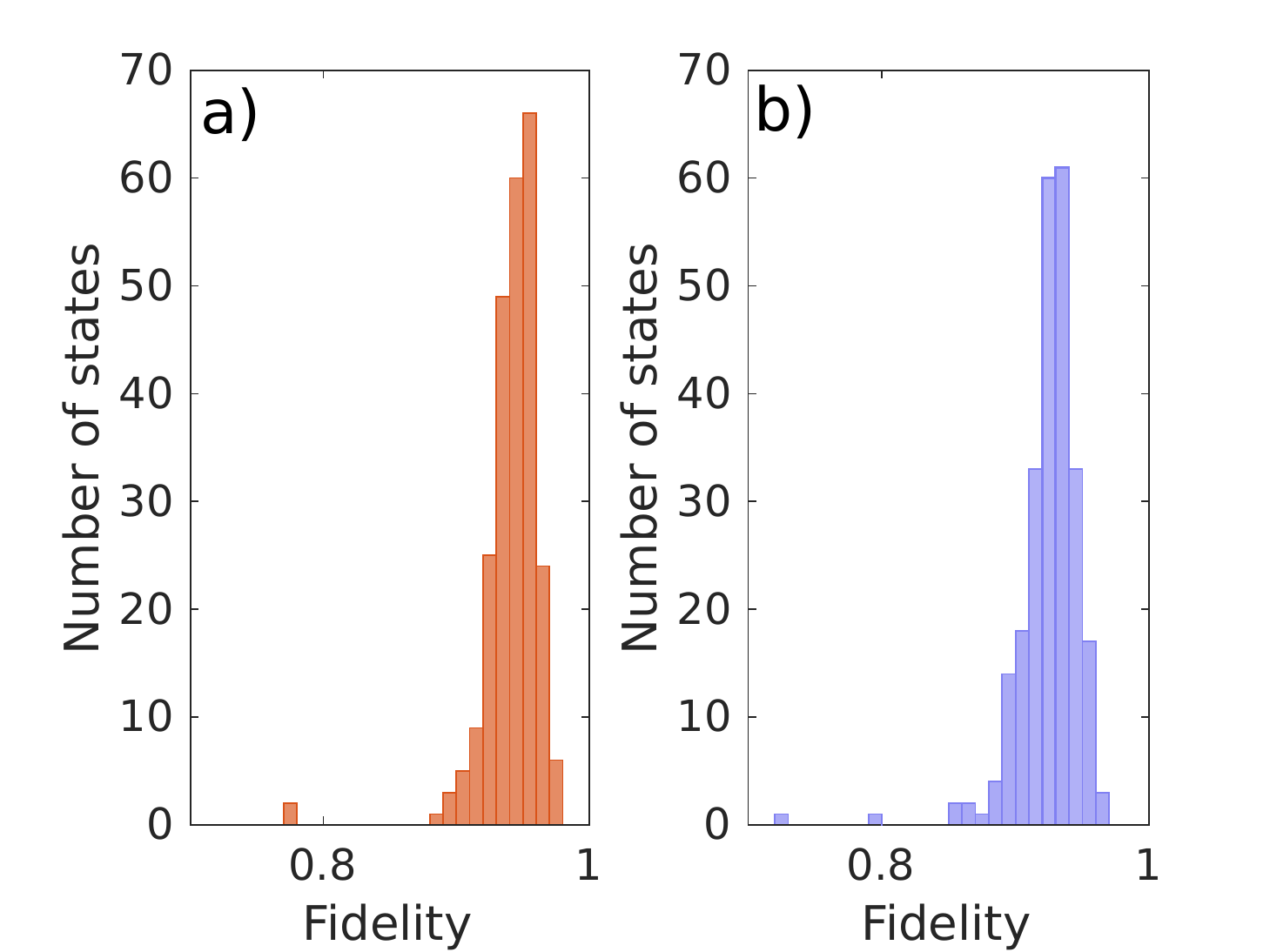}
\caption{Histogram of the state fidelity $F(\rho_{out},
\mathcal{E}_{\mathrm{exp}}\left(\rho_{in}\right))$ between the density matrix $\rho_{out}$ obtained after performing QST of a given initial state $\rho_{in}$ affected by the implemented process, and the expected density matrix $\mathcal{E}_{\mathrm{exp}}\left(\rho_{in}\right)$
corresponding to the same initial state under the action of the map $\mathcal{E}_{\mathrm{exp}}$, obtained after performing QPT of the implemented process. Obtained fidelity in the case in which standard QPT (\textbf{a}), or tensor product SEQPT (\textbf{b}), was performed. For each histogram, 250 arbitrary states of dimension $d=6$ were prepared. 
} \label{fig:histogram-exp-exp}
\end{figure}

In addition, we have analyzed how the quality in the reconstruction of
the matrix $\chi$ affects the possibility of estimating a quantum
state after the corresponding channel.  To this purpose, we performed
standard quantum state tomography (QST) for a large number of pure
states, $\rho_{in}$, randomly chosen on $\mathcal{H}$ and prepared by
$\mathrm{SLM_1}$, after being affected by the process
$\mathcal{E}_t$. We compared each reconstructed state, $\rho_{out}$,
with the predicted one by the action of the process
$\mathcal{E}_{\mathrm{exp}}$, previously obtained by means of the
SEQPT method. As figure of merit we used the fidelity between these
two states,
$F(\rho_{out}, \mathcal{E}_{\mathrm{exp}}\left(\rho_{in}\right))$. In
Fig.~\ref{fig:histogram-exp-exp}.a) we show the histogram of the
fidelity for 250 of such states. In
Fig.~\ref{fig:histogram-exp-exp}.b) we present the analogous histogram
for the case in which the process $\mathcal{E}_{\mathrm{exp}}$ was
reconstructed by means of the standard QPT method. The average state
fidelity in the case of SEQPT is
$\langle F_{\mathrm{seqpt}}\rangle = 0.925$, with a standard deviation
$\sigma_F = 0.024$, while in the case of standard QPT we obtain
$\langle F_{\mathrm{sqpt}}\rangle = 0.942$ and a similar deviation
$\sigma_F$.

\begin{figure}[t]
\includegraphics[width=.95\linewidth]{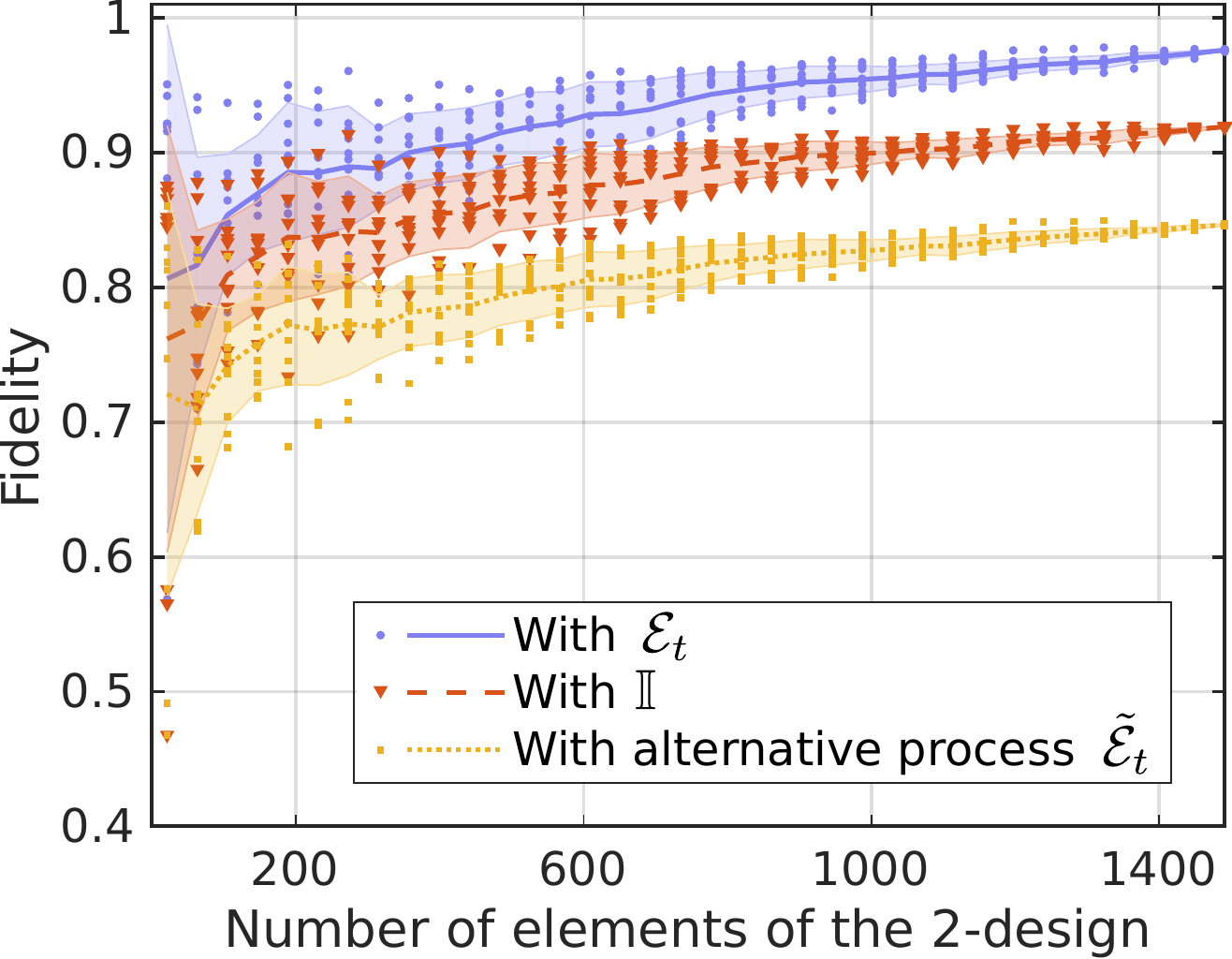}
\caption{Fidelity in the reconstruction of the implemented process for an increasing sampling on the elements of the tensor product of $2-$design. The sampling is performed only to reconstruct the 21 non-zero elements characterizing the target process
$\mathcal{E}_t$.
The different markers represent the fidelity values between the reconstructed process and different target processes $\mathcal{E}_{\mathrm{target}}$. The mean value of the fidelity is indicated by a continuous line ($\mathcal{E}_{\mathrm{target}}=\mathcal{E}_t$), a dashed line ($\mathcal{E}_{\mathrm{target}}=\mathbb{I}$), or a dotted line ($\mathcal{E}_{\mathrm{target}}=\tilde{\mathcal{E}}_t$). In each case, the shaded areas correspond to the standard deviation.\label{fig:efficiency-curve}}
\end{figure}

The main aspect of the QPT method that we study here is that it is both \emph{selective} and \emph{efficient}. The selective property makes it ideally suited to
reconstruct target processes with few non-zero matrix elements. The target process $\mathcal{E}_{t}$ that we have implemented has, in the selected basis, 21 non-zero elements over a total of 1296 elements of the matrix $\chi$. 
The efficiency property allows to estimate each element $\chi_{j_1j_2}^{i_1i_2}$ by averaging only on a subset of size $M\leq |X_{\otimes}|$. 
To test these properties experimentally we have randomly chosen different subsets of increasing size $M$, one for each non-zero coefficient $\chi_{j_1j_2}^{i_1i_2}$, from the same data set used in the reconstruction of the full matrix. Figure
\ref{fig:efficiency-curve} shows the Choi-Jamiolkowski fidelity $F(\rho_{\mathrm{theo}},\rho_{\mathrm{exp}})$ between the target process $\mathcal{E}_t$ and the reconstructed one $\mathcal{E}_{\mathrm{exp}}$, as a function of the total number of the sampled states, $21\times M$.
To analyze the effect of the sample, we 
reconstructed each of the non-zero coefficient $\chi_{j_1j_2}^{i_1i_2}$ from several random permutations of size $M$
in the set $X_{\otimes}$, which
has a total of $72$ elements. Then, each point in the graphic illustrate one particular permutation.
It is remarkable that less of 400 measurement settings were needed to
reconstruct this processes with a fidelity above $0.9$, from a total
of $21 \times 72 = 1512$ measurement settings.  We also show the
fidelity with respect to two other target process: the identity
process $\mathbb{I}$ (dashed line), and a process
$\tilde{\mathcal{E}}_t$ (dotted line) close to $\mathcal{E}_t$, which
has the same Kraus decomposition of Eq.~(\ref{Kraus-decomp}), but
corresponding to adding a constant phase shift,
$\Delta \tilde{\varphi}=\Delta \varphi+1$rad. We can conclude that
sampling only 10 elements in $X_{\otimes}$ per non-zero coefficient
$\chi_{j_1j_2}^{i_1i_2}$, is enough to differentiate $\mathcal{E}_t$
from $\tilde{\mathcal{E}}_t$, while around 50 elements where needed to
differentiate $\mathcal{E}_t$ from the identity process.

%% file: sections/conclusions.tex
\section{Conclusions}\label{sec:conclusions}

We have presented an experimental realization of the tensor product
scheme for the SEQPT protocol. This generalizes the original SEQPT
method, allowing to efficiently and selectively characterize any
quantum process in arbitrary dimension $d$. We successfully
reconstructed a physical target process in dimension $d=6$, which is
the smallest dimension for which this SEQPT extension becomes
relevant. We explicitly show how to build, experimentally, each step
of the algorithm and tested the method in a photonic platform, showing
that it has a performance comparable to that of the QPT in the same
experimental conditions.

In addition, we verified that the reconstruction can be carried out
selectively and efficiently. For that matter, we randomly sampled on
an increasing number of elements of the tensor product of $2$-designs,
to obtain the non-zero elements of the target process matrix, which
provide enough information to distinguish it from other processes. The
resulting fidelity surpass $0.9$ by sampling only a small fraction of
the total set of states.

%% file: sections/appendix.tex
\section{Bases of the operator space and MUBs}
\label{sec:appendix}

To expand the channel $\mathcal{E}$ we have chosen two basis of unitary operators acting on $\mathcal{H}_1$ and $\mathcal{H}_2$, respectively. The selected bases are the well known Sylvester's bases \cite{Sylvester, Singh2018}, which for any dimension $d$ can be written as:
\begin{equation}
  E_{n}\equiv E_{k l} = \sum_{m=0}^{d-1} \omega^{ml} \ketbra{m \oplus k}{m},
\end{equation}
where $k,l = 0,\dots, d-1$, $\omega = \exp(2 \pi i/d)$ is a root of unity and $\oplus$ is the modulo-$d$ addition. 

For the case $d=D_1=2$, the four operators are simply
\begin{equation} 
E_{00} = \mathbb{I} \: \: , \: \: E_{01} = \sigma_z \: \: , \: \: E_{10} = \sigma_x \: \: , \: \: E_{11} = i \sigma_y \, ,
\end{equation}
from where we can obtain three abelian sets of two elements each: $\{ E_{00} , E_{01} \}$, $\{ E_{00} , E_{10} \}$ and $\{ E_{00} , E_{11} \}$. The three bases that diagonalize each of these sets, i.e. the three bases of eigenvectors of the Pauli operators, not only give a complete set of MUBs for $d=2$ (and, hence, a proper $2$--design) but also have the property that the action of any of the four operators, $E_{kl}$, over any of the elements in the $2$--design, gives another element within the same MUB basis, except for a global phase. In fact, if $\ket{\psi_m^j}$ is one of the $d$ elements within the $j$-MUB, the following property is verified:  
\begin{equation}
E_{kl} \ket{\psi_m^j} = e^{i\alpha(k,l,m,j)} \ket{\psi_{m'}^j}.
\label{Op_on_MUB}
\end{equation}

In the case that $d=D_2=3$, we can analogously obtain a $2$--design by extracting four abelian subsets from the nine operators $E_{kl}$. The first of them, $\{ E_{00},E_{01},E_{02} \}$, is diagonalized by the canonical basis
\begin{equation}
\mathcal{B}_1 = \left\{ \ket{0} , \ket{1} , \ket{2}   \right\} \equiv \left\{ (1,0,0) , (0,1,0) , (0,0,1)   \right\} \, .
\end{equation}
The next set, $\{ E_{00}, E_{10} , E_{20} \}$ is diagonalized by:
\begin{equation}
\mathcal{B}_2 = \left\{  \frac{(1,1,1)}{\sqrt{3}}  ,  \frac{(1,\omega,\omega^2)}{\sqrt{3}} , \frac{(1,\omega^2,\omega)}{\sqrt{3}}  \right\} \, ,
\end{equation}
where $\omega = \exp \left(2i\pi/3 \right)$, $\omega^2 = \omega^*$ and  $\omega^3 = 1$. It is clear that $\mathcal{B}_1$ and $\mathcal{B}_2$ are mutually unbiased. Moreover, by taking $\mathcal{B}_3$ and $\mathcal{B}_4$ as the bases that diagonalize the sets $\{ E_{00} , E_{11} , E_{22} \}$ and $\{ E_{00} , E_{12} , E_{21} \}$ respectively, we get four MUBs in $d=3$ and hence a $2$--design in the corresponding Hilbert space. 
Again, it is easy to check that the property given by Eq.~(\ref{Op_on_MUB}) holds for the $9$ operators $E_{kl}$.

If the $2$--designs in each subsystem $X_1$ and $X_2$ are chosen as the complete sets of MUBs obtained above, the action of any element of the operator basis over any element of $X_\otimes$ gives, by construction, another element of $X_\otimes$. Thus, the experimental implementation of the \emph{modified} quantum channel $\mathcal{E}_{j_1j_2}^{i_1i_2}$ only requires preparing products of the two design elements as input states $\ket{\phi_A}$ (Fig.\ref{fig:circuit})
.